\documentclass[%
 aip,
 amsmath,amssymb,
reprint,%
]{revtex4-1}

\usepackage{graphicx}% Include figure files
\usepackage{dcolumn}% Align table columns on decimal point
\usepackage{bm}% bold math
\usepackage[utf8]{inputenc}
\usepackage[T1]{fontenc}
\usepackage{amsmath}
\usepackage{mathptmx}
\usepackage{etoolbox}
\usepackage{multirow}
\usepackage[hidelinks]{hyperref}
\usepackage{subcaption}
\usepackage{orcidlink}

\usepackage[final]{changes}

\makeatletter
\def\@email#1#2{%
 \endgroup
 \patchcmd{\titleblock@produce}
  {\frontmatter@RRAPformat}
  {\frontmatter@RRAPformat{\produce@RRAP{*#1\href{mailto:#2}{#2}}}\frontmatter@RRAPformat}
  {}{}
}%
\makeatother

\begin{document}

% Credit line for the arxiv submission
\onecolumngrid
%\noindent The following article has been submitted to The Journal of Chemical Physics. After it is published, it will be found at \href{https://pubs.aip.org/aip/jcp}{https://pubs.aip.org/aip/jcp}.
\noindent The following article has been accepted by The Journal of Chemical Physics. After it is published, it will be found at \hyperlink{https://doi.org/10.1063/5.0297009}{https://doi.org/10.1063/5.0297009}.
\vspace{1em}

\title{Relativistic quintuple-zeta basis sets for the s block}
\author{Marten L. Reitsma\orcidlink{0000-0002-8255-7480}}
\affiliation{Van Swinderen Institute for Particle Physics and Gravity, University of Groningen, Nijenborgh 4, 9747 AG Groningen, The Netherlands}
\author{Eifion H. Prinsen\orcidlink{0009-0008-9660-7152}}
\affiliation{Van Swinderen Institute for Particle Physics and Gravity, University of Groningen, Nijenborgh 4, 9747 AG Groningen, The Netherlands}
\author{Johan D. Polet}
\affiliation{Van Swinderen Institute for Particle Physics and Gravity, University of Groningen, Nijenborgh 4, 9747 AG Groningen, The Netherlands}
\author{Anastasia Borschevsky\orcidlink{0000-0002-6558-1921}}
%\email{a.borschevsky@rug.nl}
\affiliation{Van Swinderen Institute for Particle Physics and Gravity, University of Groningen, Nijenborgh 4, 9747 AG Groningen, The Netherlands}
\author{Kenneth G. Dyall\orcidlink{0000-0002-5682-3132}}
\affiliation{Dirac Solutions, 10527 NW Lost Park Drive, Portland, OR 97229, U.S.A.}
\email{diracsolutions@gmail.com}

\def\Eh{$E_{\rm h}$}
\def\uEh{$\mu E_{\rm h}$}
\def\ph{\phantom{0}}
\def\phm{\phantom{-}}

\begin{abstract}
Relativistic basis sets of quintuple-zeta quality are presented for the s-block elements. The basis sets include SCF exponents for the occupied spinors and for the np shell (the latter is considered here a valence orbital). Valence and core correlating functions were optimized within multireference SDCI calculations for the ground valence configuration. Diffuse functions optimized for the corresponding anions or derived from neighboring elements are also provided.
The new basis sets were applied to a range of basic atomic and molecular properties for benchmarking purposes. Smooth convergence to the basis set limit is observed with increased basis set quality from existing double-zeta, triple-zeta, and quadruple-zeta to the newly developed quintuple-zeta basis sets. Use of these basis sets in combination with state-of-the-art approaches for treatment of relativity and correlation will allow achieving higher accuracy and lower uncertainty than previously possible in calculations on heavy atoms and molecules. 
The basis sets are available at https://doi.org/10.5281/zenodo.17088050.
\end{abstract}

\maketitle

\section{Introduction}

Precision experiments on atoms and molecules are promising subjects for testing the Standard Model and for the search for the tiny signatures of physics beyond the Standard Model. Atoms and molecules can be used to probe a wide variety of physical phenomena, such as parity violation, violation of the P,T-inversion symmetry, variation of fundamental constants, and the presence of certain types of dark matter~\cite{SafBudDeM18}. Such experiments benefit from strong enhancement effects, due to the atomic or molecular electronic structure, that amplify the otherwise tiny signals and bring them into the measurable range. Heavy atoms and molecules are particularly advantageous for such experiments as sensitivity to violations of fundamental symmetries tends to scale very rapidly with the proton number, typically as $Z^2$ to $Z^5$, depending on the sought effect~\cite{SafBudDeM18,Arrowsmith-Kron_2024}. However, such experiments are also inherently challenging, due to the unprecedented sensitivity required in order to detect the vanishingly small signatures of new physics or deviations from the Standard Model predictions. Success of these ambitious experiments thus requires specially developed experimental techniques and the availability of strong and reliable theoretical support. Electronic structure theory can provide invaluable contributions in all stages of the experiment, from conception and planning to interpretation of results. In particular, extraction of various properties of interest, such as parity violation parameters, from measured energy shifts and splittings requires knowledge of atomic or molecular coupling factors. These coupling factors are unique to the given atom or molecule and to the specific electronic state and cannot be measured in principle, necessitating their accurate predictions~\cite{SafBudDeM18}. 

Such theoretical predictions should be based on accurate and reliable calculations that incorporate a high-quality treatment of both relativistic effects and electron correlation. To be useful for planning and interpreting experiments, the theoretical predictions should also have quantifiable and reliable uncertainties. State-of-the-art computational methods, such as the relativistic coupled cluster approach, now allow us to reach meV level accuracy for atomic and molecular transition energies and ionization potentials or electron affinities, and an accuracy of a single percent for a variety of properties (see, e.g., Refs.~\onlinecite{Pasteka:17,LeiKarGuo20,Skr21,HaaEliIli20}). Such accuracy requires incorporation of the higher order relativistic corrections (Breit contribution and Lamb shift) and higher excitations, beyond the standard coupled cluster singles and doubles with perturbative triples method (CCSD(T)). A balanced high-accuracy calculation also requires use of high-quality, converged basis sets. Currently, the accuracy of the calculated values is frequently limited by the quality of the available basis sets, even when extrapolating the results to the complete basis set (CBS) limit~\cite{LeiKarGuo20,GusRicRei20, KanYanBis20,GuoPasEli21,GuoBorEli22}. This shortcoming serves as a motivation for extensions and developments of higher-quality basis sets that will allow us to push the accuracy of atomic and molecular calculations even further, providing even stronger theoretical support for a variety of challenging experiments.

%\textbf{AB: this should be written properly, including citations. Motivation: we can now reach meV precision for atomic and even molecular properties, and can quantify uncertainties (cite our recent papers). We see that these are in many cases limited by the basis set quality, which is also often the greatest source of uncertainty. Thus, development of 5z level basis sets will allow further increase in accuracy and tightening of error bars. This in turn will be crucial in supporting challenging experiments on atoms and molecules.}

%\textbf{We might want to rewrite the following, as some of it has been discussed above, and it's the same as in the p block paper. Maybe summarize a bit, as we've already mentioned the limitation on the basis set quality. The first paragraph can probably be removed.}

High-quality basis sets for heavy elements must be capable of describing atoms in which relativistic effects, both scalar and spin-dependent, have a significant impact on the wave function. These basis sets must include functions for describing the SCF (self-consistent field) orbitals and also for the inclusion of electron correlation. Calculations with basis sets that are truly saturated, both with respect to SCF and correlation orbitals, would in many cases be prohibitively expensive. It is thus common to perform calculations with a series of basis sets whose size and quality increase along the series, and to extrapolate the results to the CBS limit. The correlation-consistent basis sets~\cite{cc1,cc2,cc3,cc4} provide a systematic means of carrying out such extrapolations. These basis sets have been extended into the relativistic domain with spin-free relativistic Hamiltonians, but their coverage does not include superheavy elements.

For heavy-element calculations, the Dyall basis sets~\cite{dyall_pdz,dyall_ptz,dyall_5d,dyall_pqz,dyall_4d,dyall_5f,dyall_s,dyall_5drev,dyall_4f,dyall_6d,dyall_7p,dyall_core_corr,dyall_light} are commonly used, as they cover all elements up to Z=118 at the double-zeta (2z), triple-zeta (3z), and quadruple-zeta (4z) levels. The correlation energy CBS limit can currently be obtained from extrapolation of the 3z and 4z results (as 2z basis sets are insufficient for an accurate description of correlation effects). The extrapolation commonly uses one of a few schemes, such as those presented in Refs.~\onlinecite{HelKloKoc97,doi:10.1021/acs.jctc.9b00705,Martin1996,MARTIN2006}. However, this extrapolation is often shown to be the limiting factor in the accuracy of the calculated energies and properties~\cite{LeiKarGuo20,Skr21,HaaEliIli20}.

In order to overcome this limitation and increase the accuracy even further, basis sets of quintuple-zeta (5z) quality are needed. This paper reports the development and testing of 5z quality basis sets for the s-block elements. This work continues the ongoing effort~\cite{dyall_pdz,dyall_ptz,dyall_5d,dyall_pqz,dyall_4d,dyall_5f,dyall_s,dyall_5drev,dyall_4f,dyall_6d,dyall_7p,dyall_core_corr,dyall_light} to provide high-quality basis sets for all-electron correlated relativistic calculations across the periodic table and in particular for the heaviest elements. In addition to the SCF sets, which include functions to describe the ns and np shells, the basis sets include high-angular momentum correlating functions for valence and core shells, functions for dipole polarization for the outer core shells, and diffuse functions. The performance of these basis sets is illustrated in calculations of basic atomic and molecular properties of the s-block elements and selected diatomic molecules. \added{Two of these molecules, BaF and RaF, are currently being investigated experimentally in the context of tests of the Standard Model, due to their relatively simple electronic structure and enhanced sensitivity to a variety of new physics phenomena~\cite{IsaHoeBer10, beloy2011effect, SafBudDeM18,AggBetBor18, AltAmmCah18,  VutHorHess18, KogChaBha25, GarBerBil20, AthWilSkr25}.}

\section{Methods}

The methods have been described in detail in earlier works~\cite{dyall_pdz,dyall_ptz,DyFaeBas,Sethetal}. In particular, the specific approaches used for the earlier s-block basis sets~\cite{dyall_s,dyall_sd_diffuse} were followed in the present work; these are summarized below. The basis set extensions of the GRASP program\cite{DyFaeBas} and the RAMCI program\cite{Sethetal} were used for the basis set optimizations.

\subsection{SCF sets}

The SCF basis sets are optimized in Dirac-Hartree-Fock (DHF) calculations using the Dirac-Coulomb Hamiltonian, with the standard Gaussian nuclear charge distribution~\cite{VissDy} calculated using the mass number for the most abundant or stable isotope. 

The optimization minimized the average energy of the ground state configuration, unless otherwise stated. In accordance with the previously developed basis sets, $\ell$-optimization was employed, where one set of exponents is optimized for orbitals with the same $\ell$, e.g. the p$_{1/2}$ and p$_{3/2}$ share the same exponents. Functions for the np orbital were also optimized, on the np$^k$ configuration, where $k=1,2$. First, the exponents for the outer antinode of the np orbital were optimized, then these were held fixed while the remaining p exponents were reoptimized on the ground state configuration.

As the energy gains are relatively small for increasing the basis set size from 4z to 5z, basis sets of different sizes were optimized to determine the optimal number of exponents.  The final basis sets were chosen as the sets for which there were five functions contributing significantly to the outermost maxima of the $n$s and $n$p orbitals, and also, if possible, five functions for the outermost maxima of the ($n-1$)s and ($n-1$)p orbitals. 

To increase the size of the basis set from 4z to 5z, the 4z basis sets were taken as a starting set. As a general rule, increasing the zeta level involves adding one exponent for each outermost orbital maximum except for the first (i.e. $n=\ell+1$), and then adding a number of exponents for the first, which may be any number up to the total number added to all the other maxima. The exact number added depends on the nuclear size: more added exponents are needed for the light elements where the nucleus is smaller and the inner orbitals are larger, and fewer added exponents for the heavy elements where the nucleus is larger and the inner orbitals are smaller. 

A new strategy for addition of exponents is introduced, as the previous technique of adding exponents near the nucleus and allowing the outer exponents to adjust becomes inefficient for the large basis sets being optimized here. The previous strategy worked fairly well for smaller basis sets, where the energy change due to the addition of exponents was relatively large, but for the 5z basis sets the energy changes are expected to be small. Along with the inefficiency, finding an appropriate minimum among the multiple minima on the energy surface is a challenge, as is ensuring the desired number of exponents for each antinode.

The new strategy replaces the input exponents for an antinode with a larger even-tempered set, as an initial guess for optimization. The input exponents come from the 4z basis set, and the number of exponents for an antinode is increased by 1. In the even-tempered set, the range of the exponents is expanded by 10\% with respect to the range covered by the input exponents. This small expansion of the range helps to prevent linear dependence within the exponent set for the antinode, while avoiding collisions with the adjacent antinode sets. The expansion is achieved by retaining the smallest exponent from the input set for the antinode and multiplying the largest exponent from the input set for the antinode by 1.1, then filling the space between these two exponents with an even-tempered set. The even-tempered ratio is determined from the ratio of the new largest exponent to the new smallest exponent. 

The exponents in the input basis set that are larger than the ones in the new set for the antinode (the ``interior'' exponents) are then scaled up to adjust for the new exponent range for the antinode. The scaling factor decreases linearly with the logarithm of the exponent from 1.1 for the smallest exponent of the interior set to 1.0 for the largest.

This strategy provides a new starting set of exponents, which are then optimized by energy minimization.

Due to the large size and high density of the 5z basis sets, there is a substantial degree of linear dependence of the basis functions. This affects the numerical accuracy of the optimizations, in particular the tolerance to which the SCF calculations can be converged, and thereby the tolerance to which the exponent optimization can be converged. Optimizing all the exponents simultaneously often resulted in little or no improvement, so optimization was done on groups of adjacent exponents, cycling through the entire set by groups.
\subsection{Correlating sets}

Correlating functions were determined in the style of the correlation-consistent basis sets~\cite{cc1}, where $\ell_{\text{max}}$ is increased by one for each zeta level and the number of functions for each $\ell$ is increased by one.

Valence correlating functions for the group 2 elements were optimized in multireference singles and doubles configuration interaction (MR-SDCI) calculations on the 4:1 weighted average energy of the valence ns$^2$ and np$^2$ states, as the latter configuration mixes significantly with the former due to near-degeneracy. The correlating space consisted of a set of 4s4p4d3f2g1h primitive Gaussians, properly canonicalized to produce 4-spinor functions for use in the MR-SDCI calculations. The s and p correlating functions were taken to be the first, second, fourth, and fifth outermost functions from the SCF valence set, and the rest were optimized.

For group 1 there is no correlation within the ns$^1$ configuration, and therefore MR-SDCI calculations cannot be used to generate correlating functions. Instead, a 3d2f1g set was chosen to supplement the s and p sets for the ns and np orbitals, making use of the lower cardinality basis sets as follows. The exponents for the d functions were generated from the three valence s functions from the triple-zeta basis set, multiplied by 7/3 to make the mean square radius the same as those of the s functions. Likewise, the exponents for the f functions were generated from the two valence s functions from the double-zeta set, multiplied by 9/3, and the exponent of the g function was chosen as the second function from the triple-zeta valence s set, multiplied by 11/3. These functions we refer to as ``valence flexibility'' functions rather than ``valence correlating'' functions. 

Correlating functions for the inner shells were optimized in MR-SDCI calculations in which single and double excitations into the correlating space from all orbitals in a given shell were included. In these excitations, the angular momentum coupling of the partially occupied inner shell to the correlation functions was constrained to a zero value. This reduces the length of the CI expansion for the optimization and eliminates coupling to the open valence shells, thereby restricting the correlation to the inner shell. The correlating space depends on the highest angular momentum of the orbitals in the shell: for $\ell_\mathrm{max}=0$, the correlating space is 4s4p3d2f1g; for $\ell_\mathrm{max}=1$, the correlating space is 4s4p4d3f2g1h; for $\ell_\mathrm{max}=2$, the correlating space is 4s4p4d4f3g2h1i; and for $\ell_\mathrm{max}=3$, the correlating space is 4s4p4d4f4g3h2i1k. For the optimizations, the exponents for the occupied symmetries were chosen from the SCF set as the first, second, fourth and fifth exponents from the outer anti\-nodes for the shell, and the exponents for the unoccupied symmetries were optimized while the exponents for the occupied symmetries were held fixed. For the 1s shell, the four s correlating exponents were chosen in relation to the exponent with the largest coefficient as the two adjacent exponents on the high side and the two adjacent exponents on the low side. The exponents for the higher shells were chosen to span the same mean radius range as the s selection.

In cases where the optimized exponents for a given shell overlapped with those for a higher shell, the general principle used was to replace the overlapping exponents from the lower shell with the exponents from the higher shell. The replacement exponents were held fixed while reoptimizing the exponents from the lower shell. For valence correlation of the group 2 elements, there was a significant overlap of the valence d set with the outer core d set. Here, the two largest d exponents of the valence set were replaced by the two smallest d exponents of the outer core ($n-1$) set. 

\subsection{Diffuse functions}

One diffuse s function and one diffuse p function were determined in SCF calculations on the negative ion for group 1. The s function was optimized on the s$^2$ configuration of the anion. The p function was optimized on the 4:1 weighted average of the s$^2$ and p$^2$ configurations. The diffuse d, f, and g exponents cannot be similarly optimized, due to the flatness of the energy profile for the change in exponent values. Instead, they were determined by scaling the smallest exponent from the valence flexibility set in each symmetry by a factor determined from the nearest element where optimization is possible. The factor was determined from the group 13 anion exponents~\cite{reitsma_prep_p-5z}, by taking the ratio of the diffuse exponent to the smallest correlating exponent for each symmetry. 

For the group 2 elements, the same scaling procedure was used to generate diffuse s, p, d, f, g, and h functions. For the s and p symmetries, the ratio of the diffuse s or p exponent to the smallest exponent from the neutral SCF set for group 13 was used as the scaling factor; for the higher symmetries, the ratio of the diffuse exponent to the smallest correlating exponent was used.

\subsection{Outer-core polarization functions}

Due to the fact that there are only one or two valence electrons in the s block, the polarization of the outer core by the molecular field is important in molecular calculations, where often the valence electrons are absent\cite{dyall_s}. The s and p SCF functions and the valence correlating d functions cover the range of exponents needed for polarization, but functions of higher angular momentum are needed. These polarization functions are essentially the extension of the diffuse function sets used for the negative ion in the p block. To generate functions of higher angular momentum, the following procedure was used. First, a d function was optimized to maximize the polarizability of the $(n-1)$p orbital, with the three smallest correlating d exponents for the $(n-1)$ shell included in the optimization and held fixed. The ratio of this optimized d exponent to the smallest of the $(n-1)$ shell correlating d exponents was used as a scaling factor to generate polarization exponents for each of the higher symmetries, f, g, and h. This was done by applying the scaling factor to the smallest correlating exponent for the $(n-1)$ shell in each symmetry.

\subsection{Notation}

We use a shorthand notation for different basis set compositions, where $N$z designates the cardinality, and a prefix is used to indicate the number of correlation functions that are included. The valence set (v) includes only functions correlating the outer $n$ and ($n-1$) shells, while the core-valence (cv) set also includes correlating functions for the ($n-2$) core shell. %, and all-electron (ae) set includes correlating functions for all the shells in the core. 
Basis sets that start with the letter ``a'' also include diffuse functions. For example, the basis set av5z is the 5-zeta basis set that includes diffuse functions as well as correlating functions for the $n$ and ($n-1$) shells.

\section{Basis set results}

A summary of the SCF basis sets is given in Table~\ref{basis-sizes+energies}. These do not include functions in the occupied symmetries that are added for valence flexibility or correlation. The total SCF energies for these basis sets are given %in Table~\ref{basis-energies}, 
as a reference for validation. These energies were calculated as described above,\added{ using the Gaussian nuclear charge distribution} for the specified isotope of each element, with the basis set extension of GRASP~\cite{DyFaeBas}.
\begin{table*}[ht!] 
 \caption{SCF basis set sizes for the s-block elements, and total energies in hartrees for the ground state of each element with the shown isotope number\added{, compared to numerical energies}.}\label{basis-sizes+energies}
 \centering
 \begin{tabular}{llrrcllrr}
 \noalign{\vspace{4pt}}
 \hline
 \noalign{\vspace{4pt}}
 Element & Basis & SCF \replaced{(5z)}{energy}\ph & \added{SCF (numerical)} & \hspace{1pc} & Element & Basis & SCF \replaced{(5z)}{energy}\ph & \added{SCF (numerical)} \\
 \noalign{\vspace{4pt}}
 \hline
 \noalign{\vspace{4pt}}
  $^{7}$Li & 24s14p &  $ -7.4335331$\deleted{$2$} & \added{$-7.4335331$} &&  $^{85}$Rb & 40s33p18d    & $-2979.80501$\replaced{$4$}{$372$} & \added{$-2979.805013$}\\
  $^{9}$Be & 23s12p &  $-14.57588$\replaced{$8$}{$777$} & \added{$-14.575892$} &&  $^{88}$Sr & 40s33p18d    & $-3178.0799$\replaced{$70$}{$6988$} & \added{$-3178.079969$}\\
 \noalign{\vspace{4pt}}
 $^{23}$Na & 28s18p & $-162.07808$\replaced{$7$}{$696$} & \added{$-162.078088$} && $^{133}$Cs & 43s38p23d    & $-7786.77168$\replaced{$3$}{$269$} & \added{$-7786.771668$} \\
 $^{24}$Mg & 28s18p & $-199.935066$\deleted{$34$} & \added{$-199.935067$} && $^{138}$Ba & 43s38p23d    & $-8135.645028$\deleted{$28$} & \added{$-8135.645011$} \\
 \noalign{\vspace{4pt}}
 $^{39}$K  & 35s26p & $-601.525953$\deleted{$21$} & \added{$-601.525954$} && $^{223}$Fr & 43s43p25d16f & $-24308.19431$\replaced{$2$}{$157$} & \added{$-24308.193350$} \\
 $^{40}$Ca & 35s26p & $-679.71016$\replaced{$1$}{$058$} & \added{$-679.710161$} && $^{226}$Ra & 43s43p25d16f & $-25028.18889$\replaced{$2$}{$163$} & \added{$-25028.187810$} \\
 \noalign{\vspace{4pt}}
 \hline
 \end{tabular}
\end{table*}

% Overview: exponents sharing/overlapping:
% 2s Li+Be - no
% 3s Na+MG - no
% 4s K+Ca  - no
% 5s Rb+Sr - no
% 6s Cs+Ba - 3p-4p
% 7s Fr+Ra - 3s-4s, 4s-5s, 5s-6s
%            4p-5p and 5p-6p
For the lighter elements Li--Sr five unique exponents for the outermost maximum of each orbital other than the 1s could be optimized. The same could not be done for the heavier elements Cs--Ra, because in this case the core shells can be quite compact, leading to smaller ratios between adjacent exponents than for the valence or innermost core shells. The smaller ratios can cause linear dependence problems, preventing the optimization of five unique exponents for those orbitals. There is, however, sufficient overlap between the maxima of neighboring shells that the smallest exponent of one shell can also be used as the largest exponent of the next shell. This means that some exponents can be shared between two adjacent shells. 

%%% Specifics for each set
%% 2s Li+Be
For the 2s elements it was not possible to choose SCF sets of the same size for both elements and maintain the same quality of description of the outermost antinodes. This is due to the relative difference in nuclear charge between Li and Be, which is much larger than for the other elements. This difference  causes the orbitals of Li to be more diffuse than those of Be and thus more exponents are needed for Li.

The number of functions in a given symmetry is expected to increase by at least five for a 5z basis set when going down each column, due to the additional shell. The number of exponents to describe the region near the nucleus also changes. We see this increase from the 2s block to the 3s block, as five s functions and six p functions are added. Then, from the 3s block to the 4s block, seven s functions and eight p functions are added, and another five s functions and seven p functions going from the 4s block to the 5s block. Going to the 6s block, the s set saturates near the nucleus as the nuclear size increases and the 1s orbital extent decreases. In addition, more exponents are shared. As a consequence, only three s functions and five p functions are added. This effect is even more pronounced when going from the 6s block to the 7s block. Again, five p exponents are added, but no extra s exponents are needed in this case.

\section{Applications}

To investigate the performance of the newly-developed basis sets, calculations were carried out on several atomic and molecular systems. In order to show the consistency of the new basis sets with the previous ones, we compared calculations with the 5z basis sets to those obtained with the 2z, 3z and 4z basis sets. Additionally, extrapolation to the complete basis set (CBS) limit from different quality basis sets is demonstrated.
\added{We also compare the extrapolated results to experimental data where available, for reference purposes. It should be noted, however, that these calculations do not include the higher-order effects, such as excitations beyond perturbative triples and the  Breit and QED corrections. The latter, in particular, are important for a meaningful comparison with experiment for heavy elements~\cite{LeiKarGuo20,GuoBorEli22}. It should also be noted that the basis set energy can go below the numerical energy, due to the fact that the numerical DHF energy is not a rigorous lower bound~\cite{Fae01}.}

The calculations were performed at the Dirac-Coulomb level within the SCF, CCSD and CCSD(T) methods, using the v$N$z, av$N$z, and cv$N$z basis sets. The DIRAC program~\cite{DIRAC23} was used for all of these calculations.

The CBS(4z) limit was determined from calculations using up to the quadruple-zeta basis set, while CBS(5z) also includes the quintuple-zeta basis set results in the extrapolation procedure. The SCF energies are not extrapolated, because they are converged sufficiently for the larger basis sets, i.e. $E^{\text{SCF}}_{\text{CBS(4z)}}$~=~$E^{\text{SCF}}_{\text{4z}}$ and $E^{\text{SCF}}_{\text{CBS(5z)}}$~=~$E^{\text{SCF}}_{\text{5z}}$. The correlation energies are extrapolated using the scheme proposed by Martin~\cite{Martin1996},
\begin{equation}
E_N^{\text{corr}} = E_{\text{CBS}}^{\text{corr}} + \frac{A}{N^3},
\end{equation}
where $N$ is the basis set cardinality, and the CBS limit energy $E_{\text{CBS}}^{\text{corr}}$ and fitting parameter $A$ are obtained in a least-squares fit. The 2z basis set result is not used in this extrapolation. Thus, correlation energies from calculations with the 3z and 4z basis sets are used to determine $E_{\text{CBS(4z)}}^{\text{corr}}$, while 3z, 4z and 5z calculations are used for $E_{\text{CBS(5z)}}^{\text{corr}}$.

For the CBS limit extrapolation of the properties that are presented next, the energies are always extrapolated before computing the property value. For example, the bond lengths and dissociation energies are obtained from the extrapolated potential energy curves. This means that, while the total energies decrease monotonically to the CBS limit, the properties do not necessarily exhibit the same behavior.

\subsection{Ionization potentials and electron affinities}
The ionization potentials (IP) and electron affinities (EA) are suitable atomic properties for testing the new basis sets. The energies of the neutral atom as well as the ion and the anion are required for these properties, with correlation energy varying substantially between these systems. This demands basis sets that are well-balanced, sufficiently flexible, and not biased towards any particular system.
For the IPs of group 2 elements, the cv$N$z basis sets were used, while the EAs of group 1 elements were calculated using the av$N$z basis sets. %, and included the electrons in the $n$ and $(n-1)$ orbitals in the correlated calculations.
In the coupled cluster calculations of the EAs the electrons in the $n$ and $(n-1)$ shells were correlated. In the IP calculations, the $n$, ($n-1$) and ($n-2$) electrons were correlated . A virtual space cutoff of 100~a.u. was used in all cases except for the IP of Be, where a cutoff of 200~a.u. was necessary to ensure smooth convergence to the CBS limit.

The calculated IPs and EAs are presented in Tables \ref{Group2_IPs} and \ref{Group1_EAs}. We do not present the SCF values for the EA, as in these calculations the extra electron is not bound to the atom (resulting in a negative EA), rendering these results meaningless. 

We observe smooth behaviour when going from the 4z to the 5z quality basis sets.  The CBS(4z) and CBS(5z) results differ by up to 3-4 meV for the heaviest elements. This small difference, however, can be important when meV level accuracy is desired. Furthermore, the 5z results approach the CBS limit very closely, which allows their explicit use in cases where CBS extrapolation is not practical.

\begin{table*}[ht!]
    \caption{IPs (eV) of group 2 elements as a function of basis set and correlation level. The experimental values are taken from the NIST database~\cite{NIST_ASD_S}.}
    \label{Group2_IPs}
    \begin{tabular}{llccccccl}
    \hline
    \noalign{\vspace{4pt}}
    Atom &Method  &cv2z  &cv3z  &cv4z  & cv5z& CBS(4z) & CBS(5z) &Exp.  \\ 
    \noalign{\vspace{4pt}}
    \hline
    \noalign{\vspace{4pt}}
%%% 2025-07-22 results: 
% SCF: no extrapolation; CBS(4z) = 4z, CBS(5z) = 5z
%  CC: normal extrapolation 
Be & SCF & 8.0497 & 8.0461 & 8.0458 & 8.0457 & 8.0458 & 8.0457 &  \\
 & CCSD & 9.2928 & 9.2993 & 9.3037 & 9.3047 & 9.3070 & 9.3065 &  \\
 & CCSD(T) & 9.3030 & 9.3117 & 9.3197 & 9.3210 & 9.3257 & 9.3243 & \added{9.3227} \\
 %&  &  &  &  &  &  &  & \deleted{9.3227} \\
 \noalign{\vspace{4pt}}
Mg & SCF & 6.6188 & 6.6177 & 6.6177 & 6.6177 & 6.6177 & 6.6177 &  \\
 & CCSD & 7.5716 & 7.5931 & 7.5987 & 7.5996 & 7.6027 & 7.6018 &  \\
 & CCSD(T) & 7.6055 & 7.6332 & 7.6403 & 7.6415 & 7.6455 & 7.6443 &  \added{7.6462} \\
 %&  &  &  &  &  &  &  & \deleted{7.6462} \\
 \noalign{\vspace{4pt}}
Ca & SCF & 5.1415 & 5.1407 & 5.1407 & 5.1407 & 5.1407 & 5.1407 &  \\
 & CCSD & 5.9891 & 6.0194 & 6.0265 & 6.0295 & 6.0317 & 6.0321 &  \\
 & CCSD(T) & 6.0414 & 6.0915 & 6.1031 & 6.1075 & 6.1116 & 6.1118 & \added{6.1132} \\
 %&  &  &  &  &  &  &  & \deleted{6.1132} \\
 \noalign{\vspace{4pt}}
Sr & SCF & 4.7511 & 4.7505 & 4.7505 & 4.7505 & 4.7505 & 4.7505 &  \\
 & CCSD & 5.5528 & 5.5904 & 5.5978 & 5.6030 & 5.6033 & 5.6055 &  \\
 & CCSD(T) & 5.6086 & 5.6706 & 5.6841 & 5.6916 & 5.6939 & 5.6963 & \added{5.6949} \\
 %&  &  &  &  &  &  &  & \deleted{5.6949} \\
 \noalign{\vspace{4pt}}
Ba & SCF & 4.2812 & 4.2807 & 4.2807 & 4.2807 & 4.2807 & 4.2807 &  \\
 & CCSD & 5.0399 & 5.0869 & 5.0953 & 5.0963 & 5.1014 & 5.0997 &  \\
 & CCSD(T) & 5.0990 & 5.1809 & 5.1981 & 5.2012 & 5.2106 & 5.2080 & \added{5.2117} \\
 %&  &  &  &  &  &  &  & \deleted{5.2117} \\
 \noalign{\vspace{4pt}}
Ra & SCF & 4.3597 & 4.3591 & 4.3591 & 4.3591 & 4.3591 & 4.3591 &  \\
 & CCSD & 5.0830 & 5.1521 & 5.1616 & 5.1615 & 5.1685 & 5.1655 &  \\
 & CCSD(T) & 5.1367 & 5.2443 & 5.2635 & 5.2678 & 5.2775 & 5.2753 & \added{5.2784} \\
 %&  &  &  &  &  &  &  & \deleted{5.2784} \\
    \noalign{\vspace{4pt}}
    \hline
    \end{tabular}
\end{table*}
\begin{table*}[ht!]
    \caption{EAs (eV) of group 1 elements as a function of basis set and correlation level.}\label{Group1_EAs}
    \begin{tabular}{llccccccl}
    \hline
    \noalign{\vspace{4pt}}
    Atom & Method  & av2z  & av3z  & av4z  & av5z & CBS(4z) & CBS(5z) & Exp.  \\ 
    \noalign{\vspace{4pt}}
    \hline
    \noalign{\vspace{4pt}}
%%% 2025-07-22 results: 
% SCF: no extrapolation; CBS(4z) = 4z, CBS(5z) = 5z
%  CC: normal extrapolation 
%H & SCF & -0.3303 & -0.3292 & -0.3289 & -0.3286 & -0.3289 & -0.3286 &  \\
H & CCSD & 0.7010 & 0.7376 & 0.7467 & 0.7509 & 0.7544 & 0.7547 & \added{0.75420375(3)~\cite{AndHauHot99}} \\
% & CCSD(T) & 0.7010 & 0.7376 & 0.7467 & 0.7509 & 0.7544 & 0.7547 &  \\
 %&  &  &  &  &  &  &  & \deleted{0.75420375(3)~\cite{AndHauHot99}} \\
 \noalign{\vspace{4pt}}
%Li & SCF & -0.1234 & -0.1227 & -0.1225 & -0.1224 & -0.1225 & -0.1224 &  \\
Li & CCSD & 0.6024 & 0.6062 & 0.6075 & 0.6079 & 0.6087 & 0.6082 &  \\
 & CCSD(T) & 0.6075 & 0.6134 & 0.6167 & 0.6172 & 0.6193 & 0.6184 & \added{0.618049(22)~\cite{Li_EA}} \\
 %&  &  &  &  &  &  &  & \deleted{0.618049(22)~\cite{Li_EA}} \\
 \noalign{\vspace{4pt}}
%Na & SCF & -0.1040 & -0.1032 & -0.1032 & -0.1039 & -0.1032 & -0.1039 &  \\
Na & CCSD & 0.5122 & 0.5217 & 0.5232 & 0.5237 & 0.5243 & 0.5241 &  \\
 & CCSD(T) & 0.5287 & 0.5437 & 0.5464 & 0.5471 & 0.5483 & 0.5480 & \added{0.547926(25)~\cite{Na_EA}} \\
 %&  &  &  &  &  &  &  & \deleted{0.547926(25)~\cite{Na_EA}} \\
 \noalign{\vspace{4pt}}
%K & SCF & -0.0788 & -0.0782 & -0.0782 & -0.0783 & -0.0782 & -0.0783 &  \\
K & CCSD & 0.4466 & 0.4585 & 0.4604 & 0.4610 & 0.4617 & 0.4617 &  \\
 & CCSD(T) & 0.4659 & 0.4936 & 0.4980 & 0.4994 & 0.5012 & 0.5011 & \added{0.501459(13)~\cite{K_EA}} \\
 %&  &  &  &  &  &  &  & \deleted{0.501459(13)~\cite{K_EA}} \\
 \noalign{\vspace{4pt}}
%Rb & SCF & -0.0697 & -0.0694 & -0.0695 & -0.0694 & -0.0695 & -0.0694 &  \\
Rb & CCSD & 0.4219 & 0.4383 & 0.4408 & 0.4420 & 0.4428 & 0.4430 &  \\
 & CCSD(T) & 0.4434 & 0.4764 & 0.4818 & 0.4841 & 0.4858 & 0.4861 & \added{0.485916(21)~\cite{Rb_EA}} \\
 %&  &  &  &  &  &  &  & \deleted{0.485916(21)~\cite{Rb_EA}} \\
 \noalign{\vspace{4pt}}
%Cs & SCF & -0.0603 & -0.0601 & -0.0601 & -0.0600 & -0.0601 & -0.0600 &  \\
Cs & CCSD & 0.4024 & 0.4180 & 0.4211 & 0.4219 & 0.4234 & 0.4231 &  \\
 & CCSD(T) & 0.4269 & 0.4606 & 0.4674 & 0.4696 & 0.4723 & 0.4722 & \added{0.4715983(38)~\cite{Cs_EA}} \\
 %&  &  &  &  &  &  &  & \deleted{0.4715983(38)~\cite{Cs_EA}} \\
 \noalign{\vspace{4pt}}
%Fr & SCF & -0.0470 & -0.0466 & -0.0466 & -0.0465 & -0.0466 & -0.0465 &  \\
Fr & CCSD & 0.4044 & 0.4268 & 0.4292 & 0.4324 & 0.4310 & 0.4330 &  \\
 & CCSD(T) & 0.4318 & 0.4728 & 0.4783 & 0.4846 & 0.4823 & 0.4861 & \added{--} \\
 %&  &  &  &  &  &  &  & \deleted{--} \\
    \noalign{\vspace{4pt}}
    \hline
    \end{tabular}
\end{table*}

\begin{figure}[!htbp]
    \centering
    \includegraphics[width=0.85\columnwidth]{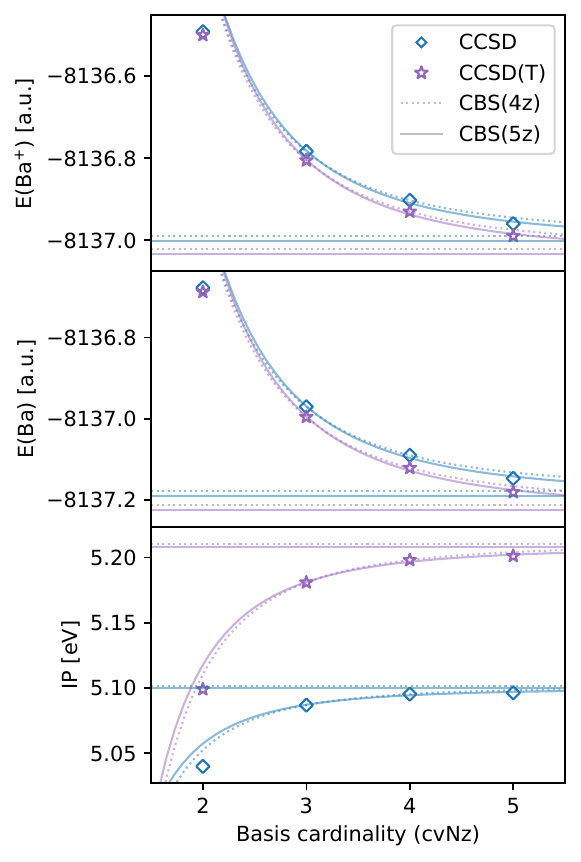}
    \caption{IP (bottom panel), ion and atom energies (top panels) of Ba as a function of basis set cardinality, calculated with CCSD and CCSD(T). Markers indicate the calculated values and the CBS(4z) and CBS(5z) extrapolations are shown as dotted and solid lines respectively.}
    \label{fig:IP+E-E_Ba_CCSD+CCSD(T)_combined}
\end{figure}

\begin{figure}[!htbp]
    \centering
    \includegraphics[width=0.85\columnwidth]{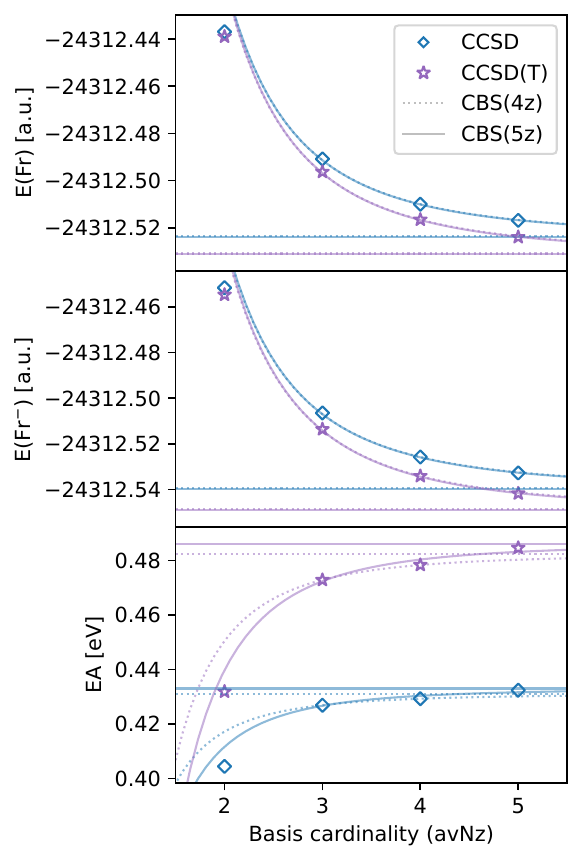}
    \caption{EA (bottom panel), atom and anion energies (top panels) of Fr as a function of basis set cardinality, calculated with CCSD and CCSD(T). Markers indicate the calculated values and the CBS(4z) and CBS(5z) extrapolations are shown as dotted and solid lines respectively.}
    \label{fig:EA+E-E_Fr_CCSD+CCSD(T)_combined}
\end{figure}

The benchmarks of the IP of Ba are visualized in Fig.~\ref{fig:IP+E-E_Ba_CCSD+CCSD(T)_combined}. This shows the smooth convergence towards the basis set limit for the energies of the ion and the atom as well as for the IP.
%MLR: Note that the extrapolation curves are of the correlation energy only, which is shown on top of the total SCF + correlation energy, by offsetting the correlation energy curve by SCF(4z) or SCF(5z). This works well, because the SCF energy is far more converged than the correlation energy, thus the full behavior is captured by the correlation CBS curve.
In a similar fashion, the EA benchmarks for Fr can be found in Fig.~\ref{fig:EA+E-E_Fr_CCSD+CCSD(T)_combined}, where the 4~meV correction to the CBS limit due to the addition of the 5z result is clearly observed, most notably for the CCSD(T) value of the EA.

All the CBS(5z) IPs and EAs are in remarkably good agreement (a few meV) with experimental values. We do not expect perfect agreement with experiment, as core-correlation, excitations beyond perturbative triples, and higher order relativistic effects are missing from these calculations. \added{In particular, excitations beyond (T) have been shown to be very important for capturing the accurate values of electron affinities~\cite{Pasteka:17,LeiKarGuo20, GuoPasEli21}.} Thus, the new basis sets provide an excellent quality basis for higher order corrections. For Fr no experimental EA is available, but our result is in excellent agreement with the relativistic Fock space coupled cluster value (0.491 eV) from Ref.~\onlinecite{LanEliIsh01}.

\subsection{Bond lengths and dissociation energies}

The equilibrium bond lengths ($r_e$) and dissociation energies ($D_e$) were calculated for alkali dimers and alkaline earth fluorides. For the latter, the 5z basis set from Ref.~\onlinecite{reitsma_prep_p-5z} was used for F. Electrons in the $n$ and ($n-1$) shells were included in the correlation space for the CCSD and CCSD(T) calculations, with a virtual space cutoff set at 100~a.u., except for $D_e$ of Rb$_2$ and Cs$_2$, where a cutoff of 15~a.u. and 10~a.u. was used respectively to decrease the computational costs.
%, except for $r_e$ and $D_e$ of Fr$_2$ and for $D_e$ of Rb$_2$, where a 50~a.u. cutoff was used instead to decrease the computational costs. 
In the calculations of the dissociation energies we have employed the counterpoise approach to correct for the basis set superposition error (BSSE)~\cite{BoyBer70}. 

\begin{figure}[!htbp]
    \centering
    \includegraphics[width=0.85\columnwidth]{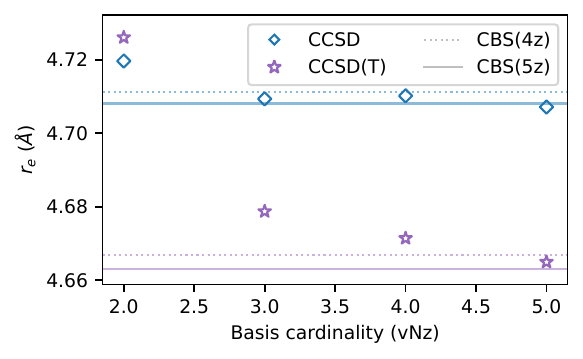}
    \caption{The bond length of Cs$_2$ using the CCSD and CCSD(T) methods as a function of basis set cardinality. Markers indicate the calculated values and the CBS(4z) and CBS(5z) extrapolations are shown as dotted and solid lines respectively.}
    \label{fig:BL_Cs2_CCSD+CCSD(T)_combined}
\end{figure}

Results for the bond lengths of the alkali dimers are shown in Table~\ref{Group1_R_e}; we excluded the Fr$_2$ dimer from this study, due to the high computational costs. %The experimental bond lengths are shown for all molecules except for Fr$_2$, for which there is no experimental bond length available; for this system bond length of 4.593 \AA and dissociation energy of 0.431 eV were predicted by a combination of scalar relativistic pseudopotentials and the CCSD(T) approach\cite{LimLaeSch02}.
%We can expect the present results to be more accurate due to a more complete treatment of relativity. 
%https://www.researchgate.net/publication/268271047_Relativistic_2-Spinor_with_Finite_Elements_Method_and_Highly_Accurate_Benchmark_Results_for_2-atomic_Molecules
%https://doi.org/10.1021/acs.jpca.6b11741
\begin{table*}[htbp!]
 \caption{Bond lengths $r_e$ (\AA) of alkali dimers as a function of basis set and correlation level.}\label{Group1_R_e}
 \begin{tabular}{llccccccl}
 \hline
 \noalign{\vspace{4pt}}
 Molecule & Method  & v2z & v3z & v4z & v5z & CBS(4z) & CBS(5z) & Exp. \\ 
 \noalign{\vspace{4pt}}
 \hline
 \noalign{\vspace{4pt}}
%%% 2025-07-22 results: 
     Na$_2$ &      SCF &   3.1923 &   3.1923 &   3.1925 &   3.1925 &   3.1925 &   3.1926 &  \\
         &     CCSD &   3.0898 &   3.0862 &   3.0841 &   3.0835 &   3.0823 &   3.0826 &  \\
         &  CCSD(T) &   3.0907 &   3.0818 &   3.0788 &   3.0780 &   3.0763 &   3.0767 & \added{3.0788(7)~\cite{RosHolGai75}} \\
 %        &          &          &          &          &          &          &          & \deleted{3.0788(7)~\cite{RosHolGai75}} \\
 \noalign{\vspace{4pt}}
      K$_2$ &      SCF &   4.1848 &   4.1848 &   4.1845 &   4.1841 &   4.1845 &   4.1841 &  \\
         &     CCSD &   3.9752 &   3.9491 &   3.9461 &   3.9446 &   3.9441 &   3.9437 &  \\
         &  CCSD(T) &   3.9858 &   3.9335 &   3.9257 &   3.9228 &   3.9202 &   3.9200 & \added{3.9051~\cite{nist-diatomic-webbook}} \\
 %        &          &          &          &          &          &          &          & \deleted{3.9051~\cite{nist-diatomic-webbook}} \\
 \noalign{\vspace{4pt}}
     Rb$_2$ &      SCF &   4.5184 &   4.5220 &   4.5218 &   4.5220 &   4.5218 &   4.5220 &  \\
         &     CCSD &   4.2521 &   4.2412 &   4.2404 &   4.2395 &   4.2402 &   4.2395 &  \\
         &  CCSD(T) &   4.2651 &   4.2230 &   4.2167 &   4.2139 &   4.2126 &   4.2119 & \added{4.2099~\cite{amiot1985laser}} \\
 %        &          &          &          &          &          &          &          & \deleted{4.2099~\cite{amiot1985laser}} \\
 \noalign{\vspace{4pt}}
     Cs$_2$ &      SCF &   5.0424 &   5.0534 &   5.0528 &   5.0533 &   5.0528 &   5.0533 &  \\
         &     CCSD &   4.7197 &   4.7094 &   4.7102 &   4.7072 &   4.7114 &   4.7081 &  \\
         &  CCSD(T) &   4.7261 &   4.6787 &   4.6714 &   4.6649 &   4.6669 &   4.6630 & \added{4.6499~\cite{honing1979high}} \\
 %        &          &          &          &          &          &          &          & \deleted{4.6499~\cite{honing1979high}} \\
 %\noalign{\vspace{4pt}}
%    Fr2 &      SCF &   4.9581 &   4.9682 &   4.9683 &   4.9680 &   4.9683 &   4.9680 &  \\
%        &     CCSD &   4.6774 &   4.6454 &   4.6535 &   4.6391 &   4.6593 &   4.6440 &  \\
%        &  CCSD(T) &   4.6888 &   4.6252 &   4.6281 &   4.6072 &   4.6302 &   4.6107 &  \\
%        &          &          &          &          &          &          &          & 4.593\footnote{Theory value from Ref.~\onlinecite{lim2005ground} obtained using CCSD(T) and a scalar relativistic pseudopotential.} \\

 \noalign{\vspace{4pt}}
 \hline
 \end{tabular}
\end{table*}
The difference between the SCF and coupled cluster results shows that electron correlation decreases the bond length for each dimer by about 0.3~\AA.
%\textbf{AB: here we should have a couple of figures and a discussion about: behaviour with respect to cardinality, extrapolation 4z vs. 5z.}
%The CBS(4z) value lies even above the 
We also note that the bond lengths decrease and get closer to the experimental values as the basis set increases to 5z and for the results extrapolated to the complete basis set limit. The CCSD(T) CBS(5z) extrapolations for all the lighter molecules in Table~\ref{Group1_R_e} perform remarkably well compared to experiment. The difference between the 4z and 5z CBS limits is very small here, which shows the reliability of the extrapolation scheme. Nevertheless, the $r_e$ of Cs$_2$ is corrected by almost 4~m\AA\  when comparing CBS(4z) and CBS(5z) at the CCSD(T) level. This is further illustrated in Fig.~\ref{fig:BL_Cs2_CCSD+CCSD(T)_combined}; while the 5z basis set does not change the CCSD result much, the CCSD(T) bond length of Cs$_2$ is lowered more significantly, demonstrating the importance of high-quality basis sets when computing higher order correlation corrections.

Table~\ref{Group1_D_e} contains the calculated dissociation energies of alkali dimers. Within the SCF treatment these systems have a wrong dissociation limit and a negative dissociation energy and thus only the coupled cluster results are presented.%, calculated using the same parameters as for $r_e$.
% https://journals-aps-org.proxy-ub.rug.nl/pra/pdf/10.1103/PhysRevA.54.R1006
\begin{table*}[htbp!]
 \caption{Dissociation energies $D_e$ (eV) of alkali dimers as a function of basis set size and correlation level.} \label{Group1_D_e}
 \begin{tabular}{llccccccl}
 \hline
 \noalign{\vspace{4pt}}
 Molecule & Method  & v2z & v3z & v4z & v5z & CBS(4z) & CBS(5z) & Exp. \\ 
 \noalign{\vspace{4pt}}
 \hline
 \noalign{\vspace{4pt}}
%%% 2025-07-22 results: 
     Na$_2$ &     CCSD &   0.5908 &   0.6794 &   0.6933 &   0.6960 &   0.7030 &   0.7012 &  \\
         &  CCSD(T) &   0.6252 &   0.7200 &   0.7365 &   0.7398 &   0.7481 &   0.7461 & \added{0.74668(1)~\cite{JonMalLet96}} \\
 %        &          &          &          &          &          &          &          & \deleted{0.74668(1)~\cite{JonMalLet96}} \\
 \noalign{\vspace{4pt}}
      K$_2$ &     CCSD &   0.3449 &   0.4431 &   0.4595 &   0.4649 &   0.4707 &   0.4707 &  \\
         &  CCSD(T) &   0.4001 &   0.5119 &   0.5338 &   0.5409 &   0.5490 &   0.5488 & \added{0.563(4)~\cite{VinVol67}} \\
 %        &          &          &          &          &          &          &          & \deleted{0.563(4)~\cite{VinVol67}} \\
 \noalign{\vspace{4pt}}
     Rb$_2$ &     CCSD &   0.2827 &   0.3775 &   0.3910 &   0.3968 &   0.4007 &   0.4017 &  \\
         &  CCSD(T) &   0.3463 &   0.4545 &   0.4737 &   0.4814 &   0.4875 &   0.4884 & \added{0.4951066(7)~\cite{SetLeRVer00}} \\
 %        &          &          &          &          &          &          &          & \deleted{0.4951066(7)~\cite{SetLeRVer00}} \\
 \noalign{\vspace{4pt}}
     Cs$_2$ &     CCSD &   0.2482 &   0.3229 &   0.3385 &   0.3427 &   0.3498 &   0.3487 &  \\
         &  CCSD(T) &   0.3156 &   0.4064 &   0.4292 &   0.4361 &   0.4458 &   0.4447 & \added{0.4525(1)~\cite{weickenmeier1985accurate}} \\
 %        &          &          &          &          &          &          &          & \deleted{0.4525(1)~\cite{weickenmeier1985accurate}} \\
%    Fr2 &     CCSD &   0.2435 &   0.2972 &   0.3062 &   0.3172 &   0.3123 &   0.3194 &  \\
%        &  CCSD(T) &   0.3166 &   0.3887 &   0.4025 &   0.4193 &   0.4122 &   0.4228 &  \\
%        &          &          &          &          &          &          &          & 0.431~\footnote{Theory value from Ref.~\onlinecite{lim2005ground} obtained using CCSD(T) and a scalar relativistic pseudopotential.} \\
\noalign{\vspace{4pt}}
 \hline
 \end{tabular}
\end{table*}
The dissociation energies obtained using the new 5z basis sets are 0.4--1.6\% higher than the 4z results; at the same time, CBS(5z) limit differs by at most 0.3\% from the CBS(4z) limit. The difference in the CBS limit of about 2~meV between the CBS(4z) and CBS(5z) extrapolations for CCSD and CCSD(T) is significant in the context of high-accuracy calculations. This demonstrates the importance and potential of these newly developed basis sets for improving the results of correlated calculations. The 5z and CBS(5z) limit results agree well with experimental values, despite the missing higher order corrections. 

%Comparison of these plots shows that the SCF calculations are already well-converged at the 4z basis set level and the 5z basis set does not noticeably affect the result, as expected. 
Finally, the bond lengths and the dissociation energies of the alkaline earth fluorides are given in Tables~\ref{Group2F_R_e} and \ref{Group2F_D_e}.
\begin{table*}[htbp!]
\caption{Bond lengths $r_e$ (\AA) of alkaline earth fluorides as a function of basis set and correlation level.} \label{Group2F_R_e}
 \begin{tabular}{llccccccl}
 \hline
 \noalign{\vspace{4pt}}
 Molecule & Method  & v2z & v3z & v4z & v5z & CBS(4z) & CBS(5z) & Exp. \\ 
 \noalign{\vspace{4pt}}
 \hline
 \noalign{\vspace{4pt}}
%%% 2025-07-22 results: 
     MgF &      SCF &   1.7416 &   1.7362 &   1.7370 &   1.7372 &   1.7370 &   1.7372 &  \\
         &     CCSD &   1.7538 &   1.7448 &   1.7456 &   1.7464 &   1.7457 &   1.7464 &  \\
         &  CCSD(T) &   1.7569 &   1.7478 &   1.7493 &   1.7505 &   1.7498 &   1.7507 & \added{1.7500~\cite{nist-diatomic-webbook}} \\
 %        &          &          &          &          &          &          &          & \deleted{1.7500~\cite{nist-diatomic-webbook}} \\
 \noalign{\vspace{4pt}}
     CaF &      SCF &   2.0183 &   1.9919 &   1.9792 &   1.9793 &   1.9792 &   1.9793 &  \\
         &     CCSD &   2.0230 &   1.9809 &   1.9610 &   1.9593 &   1.9561 &   1.9567 &  \\
         &  CCSD(T) &   2.0236 &   1.9800 &   1.9590 &   1.9574 &   1.9534 &   1.9544 & \added{1.967~\cite{nist-diatomic-webbook}} \\
 %        &          &          &          &          &          &          &          & \deleted{1.967~\cite{nist-diatomic-webbook}} \\
 \noalign{\vspace{4pt}}
     SrF &      SCF &   2.1299 &   2.1031 &   2.1016 &   2.1017 &   2.1016 &   2.1017 &  \\
         &     CCSD &   2.1312 &   2.0879 &   2.0822 &   2.0803 &   2.0792 &   2.0788 &  \\
         &  CCSD(T) &   2.1299 &   2.0861 &   2.0804 &   2.0787 &   2.0773 &   2.0772 & \added{2.07537~\cite{nist-diatomic-webbook}} \\
 %        &          &          &          &          &          &          &          & \deleted{2.07537~\cite{nist-diatomic-webbook}} \\
 \noalign{\vspace{4pt}}
     BaF &      SCF &   2.2657 &   2.2139 &   2.2072 &   2.2073 &   2.2072 &   2.2073 &  \\
         &     CCSD &   2.2654 &   2.1965 &   2.1846 &   2.1817 &   2.1808 &   2.1799 &  \\
         &  CCSD(T) &   2.2633 &   2.1945 &   2.1823 &   2.1797 &   2.1784 &   2.1777 & \added{2.162~\cite{nist-diatomic-webbook}} \\
 %        &          &          &          &          &          &          &          & \deleted{2.162~\cite{nist-diatomic-webbook}} \\
 \noalign{\vspace{4pt}}
     RaF &      SCF &   2.3443 &   2.2774 &   2.2687 &   2.2688 &   2.2687 &   2.2688 &  \\
         &     CCSD &   2.3462 &   2.2607 &   2.2468 &   2.2444 &   2.2430 &   2.2425 &  \\
         &  CCSD(T) &   2.3457 &   2.2596 &   2.2455 &   2.2433 &   2.2415 &   2.2413 & \added{--} \\
 %        &          &          &          &          &          &          &          & \deleted{--} \\
                                        %     \multicolumn{8}{c}{} &2.2357~\cite{WilPerUdr24}~\footnote{TTheory value from Ref.~\onlinecite{WilPerUdr24} obtained using relativistic CCSD(T) with higher order corrections.}  \\%bond length RaF 2.29 (theory, see https://arxiv.org/pdf/1302.5682)
\noalign{\vspace{4pt}}
\hline
\end{tabular}
\end{table*}

\begin{table*}[htbp!]
\caption{Dissociation energies (eV) of alkaline earth fluorides as a function of basis set and correlation level.}\label{Group2F_D_e}
\begin{tabular}{llccccccl}
\hline
\noalign{\vspace{4pt}}
Molecule & Method  & v2z  & v3z  & v4z  & v5z & CBS(4z) & CBS(5z) & Exp.  \\ 
\noalign{\vspace{4pt}}
\hline
\noalign{\vspace{4pt}}
%%% 2025-07-22 results: 
     MgF &      SCF &   3.4526 &   3.6289 &   3.6504 &   3.6525 &   3.6504 &   3.6525 &  \\
         &     CCSD &   3.8761 &   4.2693 &   4.4113 &   4.4598 &   4.4992 &   4.5044 &  \\
         &  CCSD(T) &   3.9423 &   4.3549 &   4.5112 &   4.5658 &   4.6095 &   4.6156 & \added{4.662(87)\cite{Eng79}} \\
 %        &          &          &          &          &          &          &          & \deleted{4.662(87)\cite{Eng79}} \\
 \noalign{\vspace{4pt}}
     CaF &      SCF &   3.9982 &   4.2602 &   4.3575 &   4.3619 &   4.3575 &   4.3619 &  \\
         &     CCSD &   4.4587 &   5.0184 &   5.3031 &   5.3699 &   5.4398 &   5.4404 &  \\
         &  CCSD(T) &   4.5097 &   5.0909 &   5.4023 &   5.4763 &   5.5585 &   5.5572 & \added{5.456(87)\cite{Eng79}} \\
 %        &          &          &          &          &          &          &          & \deleted{5.456(87)\cite{Eng79}} \\
 \noalign{\vspace{4pt}}
     SrF &      SCF &   4.0552 &   4.2882 &   4.3339 &   4.3404 &   4.3339 &   4.3404 &  \\
         &     CCSD &   4.5682 &   5.1035 &   5.3200 &   5.3945 &   5.4446 &   5.4574 &  \\
         &  CCSD(T) &   4.6277 &   5.1767 &   5.4128 &   5.4956 &   5.5518 &   5.5657 & \added{5.530(87)\cite{Eng79}} \\
 %        &          &          &          &          &          &          &          & \deleted{5.530(87)\cite{Eng79}} \\
 \noalign{\vspace{4pt}}
     BaF &      SCF &   4.2317 &   4.5515 &   4.6123 &   4.6240 &   4.6123 &   4.6240 &  \\
         &     CCSD &   4.7914 &   5.4305 &   5.6786 &   5.7721 &   5.8152 &   5.8402 &  \\
         &  CCSD(T) &   4.8555 &   5.5036 &   5.7724 &   5.8757 &   5.9242 &   5.9513 & \added{5.953(87)\cite{Eng79}} \\
 %        &          &          &          &          &          &          &          & \deleted{5.953(87)\cite{Eng79}} \\
 \noalign{\vspace{4pt}}
     RaF &      SCF &   3.7521 &   4.1471 &   4.2166 &   4.2312 &   4.2166 &   4.2312 &  \\
         &     CCSD &   4.3482 &   5.0455 &   5.3033 &   5.3995 &   5.4407 &   5.4681 &  \\
         &  CCSD(T) &   4.4069 &   5.1131 &   5.3913 &   5.4972 &   5.5436 &   5.5731 & \added{--} \\
 %        &          &          &          &          &          &          &          &  \deleted{--} \\
\noalign{\vspace{4pt}}
\hline
\end{tabular}
\end{table*}

\begin{figure}[!htbp]
    \centering
    \includegraphics[width=0.85\columnwidth]{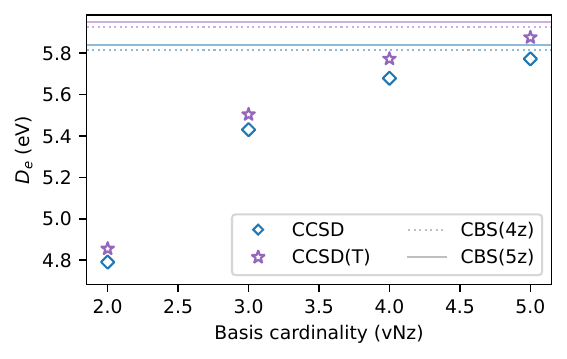}
    \caption{The dissociation energy of BaF using the CCSD and CCSD(T) methods as a function of basis set cardinality. Markers indicate the calculated values and the CBS(4z) and CBS(5z) extrapolations are shown as dotted and solid lines respectively.}
    \label{fig:DE_BaF_CCSD+CCSD(T)_combined}
\end{figure}

The basis set increase to 5z increases the bond strength, thus decreasing the equilibrium bond lengths of all the investigated molecules and increasing the dissociation energy. For BaF, the basis set convergence and CBS limits of $D_e$ are shown in Fig.~\ref{fig:DE_BaF_CCSD+CCSD(T)_combined}. Compared to the 2z and 3z results, 4z and 5z (and CBS(4z) and CBS(5z)) appear to be quite well converged. However, at the CCSD(T) level, the 4z and 5z dissociation energies differ by 103~meV, and their respective CBS limits by 27~meV, which clearly proves the need for 5z basis sets to be able to approach high accuracy.
%and we see a similar-sized contribution from the CBS(5z) extrapolation.
The bond lengths for the lighter molecules in particular agree very well with experiment; for RaF no experiment is available, but the present result is in good agreement with the recent relativistic CCSD(T) value (2.2357~\AA) from Ref.~\onlinecite{WilPerUdr24}. The dissociation energies of these molecules were measured from the reaction of metastable group IIA metals with fluorine~\cite{Eng79}; our results are in agreement with these values within the assigned experimental uncertainty of almost 100 meV. A more accurate measurements of these properties would allow for more meaningful benchmarking of computational methods for these systems. Our results for the dissociation energy of RaF are also in good agreement with the earlier 4c-CCSD(T) value of 5.54(5)~eV~\cite{WilPerUdr24}.\\

%We do note a difference between the behavior of the bond lengths here and those of the alkali dimers discussed previously. In this case the difference between SCF and CCSD results is much smaller, which is due to the much weaker bond for the alkali dimers, the shallower potential curve is much more sensitive to a change in (correlation) energy, thus the bond length is affected more than for the fluorides, which have a much higher dissociation energy. 
For the alkaline earth fluorides, correlation has a smaller effect on the calculated bond lengths and dissociation energies than for the alkali dimers. The latter systems exhibit weaker covalent bonding, strongly affected by correlation effects, while the stronger bond of ionic character in alkaline earth metal fluorides is less affected by correlation.

\section{Conclusions}
Relativistic basis sets of quintuple-zeta quality have been presented for the s-block elements. These basis sets include SCF exponents for the occupied spinors and for the np shell, which we considered as a valence orbital. Valence and core correlating functions were optimized within multireference SDCI calculations for the ground valence configuration. Diffuse functions for the corresponding anions are also provided, which are either optimized or derived from the optimized diffuse functions of neighboring elements.

The new basis sets have been benchmarked for basic spectroscopic properties for a range of atoms and molecules. The CBS limit extrapolation is demonstrated, and smooth convergence is observed with increased basis set quality from existing double-zeta, triple-zeta, and quadruple-zeta to the newly developed quintuple-zeta basis sets. The contribution of the 5z basis sets to the tested properties is shown to be at most a few percent (for $D_e$ and EA) and as small as $0.1\%$ (for $R_e$ and IP). The contribution of the 5z basis sets to the extrapolated values of these properties is even smaller.  These results give a rough estimate of the improvements that can be achieved by the use of the new 5z basis sets. In combination with state-of-the-art approaches for treatment of relativity and electron correlation, the new basis sets will allow the achievement of higher accuracy and lower uncertainty than was previously possible in calculations on heavy atoms and molecules.

The basis sets are available at https://doi.org/10.5281/zenodo.17088050.

\begin{acknowledgments}
Dutch Research Council (NWO) project number Vi.Vidi.192.088 and LISA: European Union’s H2020 Framework Programme under grant agreement no. 861198. We thank the Center for Information Technology of the University of Groningen for their support and for providing access to the Hábrók high performance computing cluster.
\end{acknowledgments}

\section*{Author declarations}
\subsection*{Conflict of Interest}
The authors have no conflicts to disclose.
\subsection*{Author Contributions}
\noindent
\textbf{Marten L. Reitsma}: Investigation; Methodology; Visualization; Writing – original draft; Writing – review \& editing.
\textbf{Eifion H. Prinsen}: Investigation; Writing – original draft; Writing – review \& editing.
\textbf{Johan D. Polet}: Investigation; Writing – review \& editing.
\textbf{Anastasia Borschevsky}: Funding acquisition; Supervision; Methodology; Writing – original draft; Writing – review \& editing.
\textbf{Kenneth G. Dyall}: Investigation; Methodology; Software; Supervision; Writing – original draft; Writing – review \& editing.

\subsection*{DATA AVAILABILITY}
The data that support the findings of this study are available from the corresponding author upon reasonable request.

\section*{References}
\bibliography{cc_basis_refs,dyall_basis_refs,dyall_refs,basis_sblock,more_refs}

\end{document}